\def\mykeywords{animation, soft-bodies, rhythmic, internal, organs, medical, oscillatory, real-time, biomechanics, computer graphics, tissue, autonomous, motions, self-driven, physics-based, tunable, customizable}

\def\mytitle{Lightweight Self-Driven Deformable Organ Animations}
\def\mysubtitle{Autonomous Procedural Rhythmic Tissue} %  Motions}

\documentclass[10pt, conference, compsocconf]{IEEEtran}
\usepackage[absolute]{textpos} % textblock

\usepackage{algorithm}
\usepackage{algpseudocode}

\usepackage{amsmath}
\usepackage{graphicx}

\usepackage{epstopdf}

\usepackage{bbm}
\usepackage{amsthm} 
\usepackage{amssymb}

\usepackage{algpseudocode}
\usepackage{algorithm}

\usepackage[backref=page]{hyperref}

\hypersetup{
    colorlinks = true, 
    linkcolor = blue,
    anchorcolor = red,
    citecolor = blue, 
    filecolor = red, 
    pagecolor = red,
    urlcolor = red,
}

\usepackage{amsfonts, amsbsy, amssymb}

\usepackage[numbers]{natbib}

\usepackage{graphicx}

\graphicspath{{./images/}}

\newcommand{\figuremacroW}[4]{
	\begin{figure}[t] %[htbp]
		\centering
		\includegraphics[width=#4\columnwidth]{#1}
		\caption[#2]{\textbf{#2} - #3}
		\label{fig:#1}
	\end{figure}
}

\newcommand{\figuremacroWb}[4]{
	\begin{figure}[b] %[htbp]
		\centering
		\vspace{-15pt}
		\includegraphics[width=#4\columnwidth]{#1}
		\caption[#2]{\textbf{#2} - #3}
		\label{fig:#1}
	\end{figure}
}

\newcommand{\figuremacroF}[4]{
	\begin{figure*}[t] % [htbp]
		\centering
		\includegraphics[width=#4\textwidth]{#1}
		\caption[#2]{\textbf{#2} - #3}
		\label{fig:#1}
	\end{figure*}
}

\let\subparagraph\relax % You don't want to use \subparagraph

\usepackage{titlesec}

\titleformat{\paragraph}[runin]
{\normalfont\normalsize\bfseries}{\theparagraph}{}{}
\titlespacing*{\paragraph}{0pt}{10pt}{5pt}

\begin{document}
\title{ \mytitle \\ \fontsize{14}{20}\selectfont \mysubtitle \vspace{-10pt} }

 \begin{textblock}{100.0}(1.5,0.8) % {block width} (coords) 
Technical Article August 2016. Lightweight Self-Driven Deformable Organ Animations
 \end{textblock}


\iffalse

\author{
	Anonymous Submission \\
	Contribution ID 15006 \\
	Anonymous Organisation
}

\fi

\iftrue

\author{
  \IEEEauthorblockN{Ben Kenwright}
  \IEEEauthorblockA{   Southampton Solent University\\
  					   School of Media Arts and Technology\\
					   United Kingdom\\
					   benjamin.kenwright@solent.ac.uk}
\iftrue
					\and
  \IEEEauthorblockN{Kanida Sinmai}
  \IEEEauthorblockA{   
  	Thaksin University\\
  	Department of Computer and Information Technology\\
  	Thailand\\
  	kanida@tsu.ac.th}
\fi
}

\fi

\maketitle

\raggedbottom

\sloppy

\iffalse
\section{abstract}
\fi

\begin{abstract}
The subject of simulating internal organs is a valuable and important topic of research to multiple fields from medical analysis to education and training. %of this kind has become a hot pot. 
%	
%The simulation of organs and issues is a valuable and important topic of research. % of controlled oscillating organs. 
This paper presents a solution that utilizes a graphical technique in combination with a Stochastic method for tuning an active physics-based model.
We generate responsive interactive organ animations with regional properties (i.e., areas of the model oscillating with different harmonic frequencies) to reproduce and capture real-world characteristics. 
%
% Our technique demonstrates the value and importance of controlled oscillating organs. 
%The research on simulating internal organs of this kind has become a hot pot. 
%
Our method builds upon biological and physical discoveries to procedurally generate internally controlled rhythmic motions but also enable the solution to be interactive and adaptive. %  interactive  the deformable organ but also the ability to internal organ through rhythmic oscillations. %  are synthesized. 
We briefly review deformation models for medical simulations and investigate the impediments to combining `computer-graphics' representations with biomechanical models.
Finally, we present a lightweight solution that is scalable and able to procedurally generate large organ animations.  In particular, simplified geometric representations of deformable structures that use periodic coupled forces to drive themselves.

% driven by 
%deformation, tissue cutting, and force-feedback interaction.
%Last, we inspect the potential of medical simulation under the
%development of this key technology.

\end{abstract}

\begin{IEEEkeywords}
\textbf{\emph{\mykeywords}}
\end{IEEEkeywords}

% \fi
%\IEEEpeerreviewmaketitle

\section{Introduction}

%1. What is the problem?
%2. Why is it interesting and important?
%3. Why is it hard? (E.g., why do naive approaches fail?)
%4. Why hasn't it been solved before? (Or, what's wrong with previous proposed solutions? How does mine differ?)
%5. What are the key components of my approach and results? Also include any specific limitations.

%%%%%
%
%1. What is the problem?
%
%%%%%
\paragraph{Simulating Large Internal Organs}  %How can we simulate internal organs?}
A virtual system for visualizing the properties of large internal organs (e.g., lungs, heart, and kidneys) would facilitate a safe and efficient training tool.  
This tool would allow surgical trainees a virtual medium to visualize and repeatedly perform surgical procedures on a wide variety of cases. 
For instance, the influence of external disturbances during surgical procedures on the organ's elastic tissue properties or oscillatory signal patterns. %  impulses.%   on organs. %  drive muscles. %  organs.  
%in  real-time 3D for medical education and patient communication tool, featuring customizability of anatomical structure and motion pattern.  
%
%
Hence, a simulation solution that is able to capture these physical-properties while being interactive and dynamic would be valuable and important.
Not to mention, the ability to customize the virtual simulation to different situations (e.g., sizes and signal patterns).
Of course, the creation of plausible tissue deformation animations requires the combination of a variety of techniques (e.g., physical-laws and numerical simulation concepts).
While biomechanical research has developed detailed mathematical models to produce large amounts of experimental data that accurately representing these soft tissue deformations, we on the other hand follow a lightweight graphical method, for real-time computation of the animations by simplifying the structure based on low-dimensional approximations (e.g., ignoring intricate internal structures and vessels, blood flow and pressure, twisting and stretching of internal flesh) \cite{delingette1998toward,cotin1999real}.

%%%%%
%
%2. Why is it interesting and important?
%
%%%%%
%\paragraph{Why is the simulation of internal organs important?}
%The simulation of organs provides surgeons and researchers with a safer option for training and experimentation.
%Surgical simulators provide researchers and medical students access to safer options for training and experimentation.
% surgeons in minimally invasive surgery. 
%One particularly important area of research is that of deformable virtual organs and soft tissue. %  for use in haptic surgical simulators. 
%

%

%%%%%%%
%
%3. Why is it hard? (E.g., why do naive approaches fail?)
%
%%%%%%%
%\paragraph{Why is it challenging to simulate internal organs?}
\paragraph{Physical Complexity}
A low-dimensional model reduces the computational bottleneck while allowing the reproduction of correlating data. % while reducing the 
As the body's internal organs (e.g., heart and lungs) are complex systems (that perform a large number of activities) and would be difficult, if not impossible, to simulate in all their glory. % every single detail). % replicate.
For example, our heart is both incredibly powerful and delicate, the muscles in a typical human heart are able to create enough energy to drive a truck for 20 miles and keep beating even if it is separated from the body because it has its own electrical impulse signals (see Figure \ref{fig:heartoutside}).
%
%
%while the average adult heart beats 72 times a minute; 100,000 times a day; 3,600,000 a year.
%
These internal organs are driven by an assortment of physical-properties, such as, conservation of energy, coupled interaction, and fluid mechanics (flow and pressure).
Not to mention, the organs have internal muscles that contract and expand in rhythmic oscillatory patterns (coordinated synergistic motion to achieve optimal performance while conserving energy).
To capture these life-like characteristics while being physically correct and interactive is an interesting and challenging problem that we address by focusing on a specific component (i.e., animation).
%

%%%%%%%%
%
%4. Why hasn't it been solved before? (Or, what's wrong with previous proposed solutions? How does mine differ?)
%
%%%%%%%%
%\paragraph{What solutions are available for simulating internal organs?}
\paragraph{Deformable Models}
The muscles and organs in our body are soft-body mechanisms. %that obey the laws of physics.
This is made easier, knowing the fact that, during the past decade many researchers have contributed to the forefront of deformable 3D objects research \cite{BKKS,terzopoulos1987elastically} and cloth animation systems \cite{baraff1998large,kenwright2011real}.
These techniques originally developed in other areas have been exploited in the medical and biological field - enabling the synthesize of physical attributes seen in real-world flesh and muscle. %that has led into medical/biological research.
While a number of these techniques are simplifications, they possess correlating factors \cite{cotin1999real,yuan2010real} to help support their viability. %  prove their viabilityy.
%
%, see D. Terzopoulos, et.al. in [1,2], and in
%cloth animation systems, Baraff and Witkin in [4], and N. Thalmann, et.al, in [5]. 
%
%This is shown by the fact that 
With this in mind, we broadly classify current surgical simulators into either physically based models \cite{basdogan1997force,cotin1999real} or finite element models (FEM) \cite{bro1996real,ferrari2015augmented}. %  [9,10,11]. 
While many techniques have been proposed for deformable object animations \cite{gibson1997survey,delingette1998toward}, few if any combine deformable tissues with dynamic self-driven solutions (e.g., procedural generated rhythmic oscillatory signals) as we do in this paper.

%%%%%%%%%%%%
%
%5. What are the key components of my approach and results? Also include any specific limitations.
%
%%%%%%%%%%%%

% some interesting facts about organs
% statistics numbers about heart/lungs
% difference between two lungs (change with running/walking/age/health)
% how grow evolve/react to physical stimulae

% lung cancer and the rythmic signals

% provide the performance necessary for
%real-time applications. The fundamental trade off therefore is accuracy vs interactivity. In
%haptic surgical simulation this problem is especially acute, as the user must feel contact forces
%(and see the graphical deformations) that are accurate yet computed in real time. 

\paragraph{Coordinated Motions}
The overall objective is not to simulate every single characteristic of muscle tissue accurately and realistic, but to simulate the organic `motions' through coordinated control of the internal constraints (analogous to muscles contracting and expanding). %  of these organs.
This paper builds upon the work presented by Kenwright and Sinmai \cite{BKKS}, however, instead of focused on animated creatures, we simulate and control the internal frequencies/movement of regional muscle tissues (for organs), such as, the heart, kidneys, and lungs.
For example, our lightweight real-time method paves the way for medical education and patient communication tools, featuring customizability of anatomical structures and motion patterns.
Our approach addresses important issues concerned with the underlying graphical and physical models designed for surgical training simulations, as well as issues related to the real-time interactivity with, and manipulation of, these models. 
%%
%The specific application of interest is laparoscopic surgery, which is performed using endoscopes that present a video image of the organs to the clinicians. 
%
%The surgeon then performs the surgery while looking at the video monitor. 
%
%The particular focus is gall bladder surgery, which involves various gastrointestinal organs. 
%The overall objective is to simulate this environment by creating realistic, manipulable models of these organs. 
%The models are interactively manipulable and exhibit behavior both visually acceptable and physically accurate. The approach is based on the notion of active surfaces. The rationale, mathematical formalism, and visualization techniques encompassed by the methodology are described. Recent results obtained from applying these methods to the problem of endoscopic gall bladder surgery simulation are presented.<>

\paragraph{Contribution}
The key contributions of this paper are:
(1) a lightweight simulation and Stochastic tuning of internal soft-body organ motions for visualization and training;
(2) scalable particle partitioning configuration (i.e., trade accuracy for performance).

\figuremacroF
{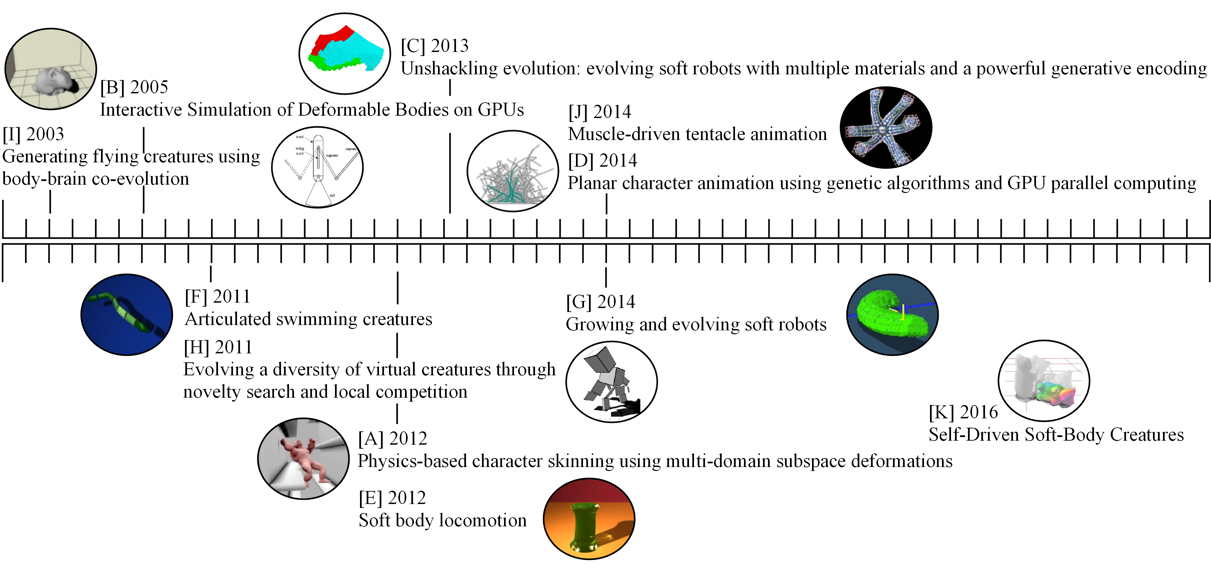}
{Timeline}
{Visual illustration of graphical and interactive related publications over the past few years that have contributed towards more active and engaging soft-body systems.
	[A] \cite{kim2012physics},
	[B] \cite{georgii2005interactive},
	[C] \cite{cheney2013unshackling},
	[D] \cite{kenwright2014planar},
	[E] \cite{tan2012soft},
	[F] \cite{tan2011articulated},
	[G] \cite{rieffel2014growing},
	[H] \cite{lehman2011evolving},
	[I] \cite{shim2003generating},
	[J] \cite{stavness2014muscle},
	[K] \cite{BKKS}.
}
{1.0}

\paragraph{Roadmap}
The structure for rest of the paper is organized as follows.  
Firstly, we briefly review existing work in Section \ref{sec:related_work}.  
Then in Section \ref{sec:method}, we present our method.
Finally, we present results in Section \ref{sec:results}, then close with a conclusion and discussion in Section \ref{sec:conclusion_and_discussion}.

%Section \ref{sec:articulated_model} describes the articulated model, we use for our simulations.  
%Then in Section \ref{sec:math_background}, we present essential mathematical algorithms and principles for the paper (e.g., dual-quaternion algebra).  
%We follow on by explaining the IK problem in Section \ref{sec:fk_ik}. While in Section \ref{sec:jacobian_matrix}, we explain the Jacobian matrix, then in Section \ref{sec:gauss_seidel_algorithm} we discuss our approach for solving the IK problem with the Gauss-Seidel algorithm.  
%Finally, we present results in Section \ref{sec:experimental_results}, then Section \ref{sec:limitations} discusses limitations, followed by the closing conclusion and discussion in Section \ref{sec:conclusion_and_discussion}.

%\begin{figure}
%	\begin{center}
%	\includegraphics[width=0.9\columnwidth]{../images/understanding_ik}
%	\end{center}
%	\caption{\text{Forward \& Inverse Kinematics}.  Illustrating the relationship between forward and inverse kinematics parameters.}
%	\label{fig:understanding_ik}
%\end{figure}

\section{Related Work} \label{sec:related_work}
Soft-body systems are a popular topic across numerous disciplines (e.g., graphical animation, robotics, and biomechanics).  
The mechanics of a passive soft-body systems are well understood and can be implemented using physical-laws (classical mechanics).
However, we are concerned with soft-body systems that are able to drive themselves.
These self-driven models possess life-like characteristics based on the physiology and biological structure.
%
%A passive is a vital component that can be implemented using a wide range of solutions.  
We give a brief overview of existing, current, and cutting-edge approaches in this area to help emphasise the different solutions the problem (enabling the reader to see where our method sits) - see Figure \ref{fig:timeline} and Figure \ref{fig:types}.

\paragraph{Biomechanics \& Physical Laws}
The study of organic soft-body tissue deformation and motion belongs to the field of biomechanics.
Physical laws are an important concept for modelling organic soft-body tissue. 
The challenges stem from the ability to accurately simulate and reproduce the deformability of real-world tissue. 
This includes, the control and tuning of deformation parameters.
% compare the computed soft-tissue model with
% the actually deformed tissue. 
The viability and accuracy of the model and simulation depends on the application; for example, surgical training requires highly realistic visual and haptic representations. %  accuracy of deformation.
\textbf{Importantly, if the simulation is too difference from the real-world behavior, it could result in learning inappropriate procedures or providing inaccurate and implausible representations}.
As we show in this paper, it is vital to have a customizable model that is able to be trained and automatically adapt based on pre-defined constraints and qualities (e.g., motion patterns, biomechanical knowledge of the organ and soft tissue properties).  This helps develop interactive solutions, since the model is able to react and solve unforeseen situations in a realistic physically plausible and life-like manner (interacting with changing surrounding).
%
%The study of soft-body tissue deformation and motion belongs to the field of biomechanics. i
% study of soft-tissue deformability belongs to the field
%of biomechanics. 
Having said that, there exists a large repertoire of information in the biomechanics literature on the study of various soft-tissue deformation properties \cite{fung2013biomechanics}. 
%Those studies range from the determination of qualitative behavior of tissues
% to the recovery of quantitative parameters governing their
%deformation. 
For example, there are specific studies on skin \cite{larrabee1986finite}, vessels \cite{fung2013biomechanics}, heart \cite{hunter1988analysis}, muscles \cite{waters1992physical}, and brain \cite{chinzei1997compression}.
Typically, these studies present static data of the tissue (e.g., elasticity, stress/strain relationship, and plasticity (non-reversible
elastic behavior)) and do not address the dynamic self-driven internal forces/motion patterns that sit within the tissue.

% static reversible elastic deformation corresponds to the linear elastic model.

\figuremacroW
{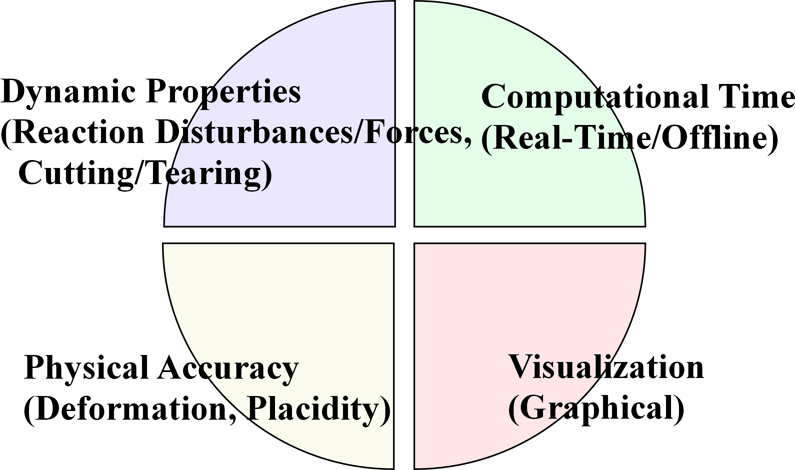}
{Categorising}
{Typical areas of investigation for achieving realistic soft-tissue models (static and dynamic).}
{0.8}

%the biomechanical behavior of soft tissue. We distinguish
%between knowledge of soft-tissue deformable properties and
%knowledge of interaction with surrounding tissues.

\paragraph{Surgical}
A number of techniques have been presented to simulate the interactive properties of tissue and organs in a medical surgical environment \cite{cover1993interactively,webster2002elastically}.
However, these techniques typically focus on static models (i.e., the underlying anatomy of the organ is either static or is animated via fixed key-frame animations).
Our work builds upon these techniques but incorporates a real-time physics-based model that is able to be tuned to create coordinated oscillatory motion signals that mimic the real-world (i.e., the rhythmic harmonic expansion and contraction of internal muscles within the organ).
We use a low-dimensional approach based on a particle and voxel partitioning system - utilizing a coupled mass-spring damper configuration (see Kenwright \& Sinmai \cite{BKKS} for further details).

\paragraph{Education Potential}
Animated model have and are a popular substitute for visualizing and teaching medical anatomy \cite{vernon2002benefits} 
(proving an invaluable teaching aid).
The advantages of a physically-based virtual solution are, 
   (1) does not decay, 
   (2) can be examined at user's leisure, 
   (3) able to interactive with,
   (4) published electronically (around world),
   and
   (5) visualize properties/characteristics of the anatomy that would be difficult to see in real-life.
There have been limitations to the capabilities and costs of creating life-like virtual models (not just the graphical detail but the physical movement). 
As computer graphics and simulation technologies have evolved, these virtual solutions have become ever more realistic (almost indistinguishable from the real-world). %replicating the real-world ).
This allows trainees to visualize/simulate the examination/surgery without the danger to patients (e.g., haptic interactive feedback interfaces) \cite{brenton2007using}.
Ultimately, these technologies visualize the subject as never seen before, offering a means to see anatomy that could never be captured by tradition film.
Our work builds upon this valuable fact, focusing on the motion characteristics of the organs (tunable customizable solution driven by a physics-based model in real-time).

% have long been used as a substitute for real
%anatomy in medical teaching. They do not decay, can be
%examined at leisure, and have proved to be valuable
%teaching aids. Virtual models, although expensive to
%produce, have built on these virtues by being easy and cheap
%to share and publish electronically. 
%
%They also have the
%potential to allow students to have simulated experiences of
%surgery without any danger to patients, and through haptic
%interfaces, can give students tactile feedback. The use of 3D
%modelling and animation also allows the visualization of
%subjects and scenes that could never be captured on film.

\paragraph{Our Work}
The ability to interact with moving organs presents an important real-time 3D experience. 
Typically, virtual anatomical solutions use pre-recorded animations which making it difficult to customize or change on-the-fly (i.e., interact with in real-time).
We solve this by integrating a lightweight model with a low-dimensional structure to emulate the physical properties and influences (interactive motions) which we are able to explore.
% the anatomical features. 
This solution enables the creation of animated transparent structures for early exploration and development (e.g., medical visualization/training). % of  the exploration of the bronchi and related key vascular components. Bronchopulmonary segments can be highlighted and labeled on the lungs and the bronchopulmonary tree.
We demonstrate and investigate preliminary proof of concept tests - plausibility of the technique.
As we see with recent trends over the past few decades, animated solutions are moving towards more intelligent self-adapting ideologies compared to static pre-recorded methods \cite{ftc2016kenwright}.

\section{Method} \label{sec:method}

A passive soft-body solution could be driven and tuned through the injection of `external' forces.
However, in reality, the organs are self-driven through the expansion and contraction of internal muscles. %) plausible using heuristic algorithms.
These internal motions typically follow rhythmic oscillatory patterns which can be reproduced using trigonometric training functions (e.g., any signal can be composed of sinusoidal signals - the concept behind Fourier series \cite{attinger1966use,FourierKenwright15}).  
Of course, this opens the door to a vast array of parameters with a finite search range that is difficult to solve in a viable time frame, while allowing real-time customization control.
% requires human intervention to achieve a desired solution.
Alternatively, we exploit a smarter solution based on a hybrid combination of methods, such as, pre-recorded training data to steer the system towards approximate solutions, in combination with human intervention to create the self-driven soft-body internal organs. %  motion that accomplishes a specific task.
We take animation data (i.e., pre-recorded key-frames).
We connect the soft-body to a training mesh via distance constraints, so 
as the animation data updates the connected particles are move in a correlating pattern.
The creation of soft-body animations are directed towards motions which mimic the desired pattern (rather than just randomly jiggling around).
This kinematic coupled solution provides a starting set of oscillating distance constraints from which we can calculate the penalty forces and inverse dynamics.
% for the point-masses and distance constraints.
When we play back the soft-body motion using the trained constraint force signals, the motions will be physically driven, and possess `approximate' features, providing a starting essence for the tuning system (evolving stochastic model).  
However, the motion will drift away from the pre-defined solution, set out by the key-frame training data (the animation will have numerical errors and external force interactions). 
This is where we need to adapt and adjust the forces to steer and control the final motions.
Allowing the user to control the final animation pattern by either forcing the system to constantly match the original training data or steer/develop new animation types based on experience (see Figure \ref{fig:overview}). 
This allows the rapid proto-typing and development of training/analysis of data, as the final simulations are generated using physics-based concepts (deformations that are organic with directable purpose).
%
%\vspace{-5pt}
Finally, due to the massive computational complexity of calculating the inter-connecting forces of a mesh model - we enable the problem to be decomposed into regions to help emphasis greater or less detail for different areas of the model.  With the internal graphical mesh updated from the physical regional mesh (see Figure \ref{fig:voxelregions}).

\figuremacroW
{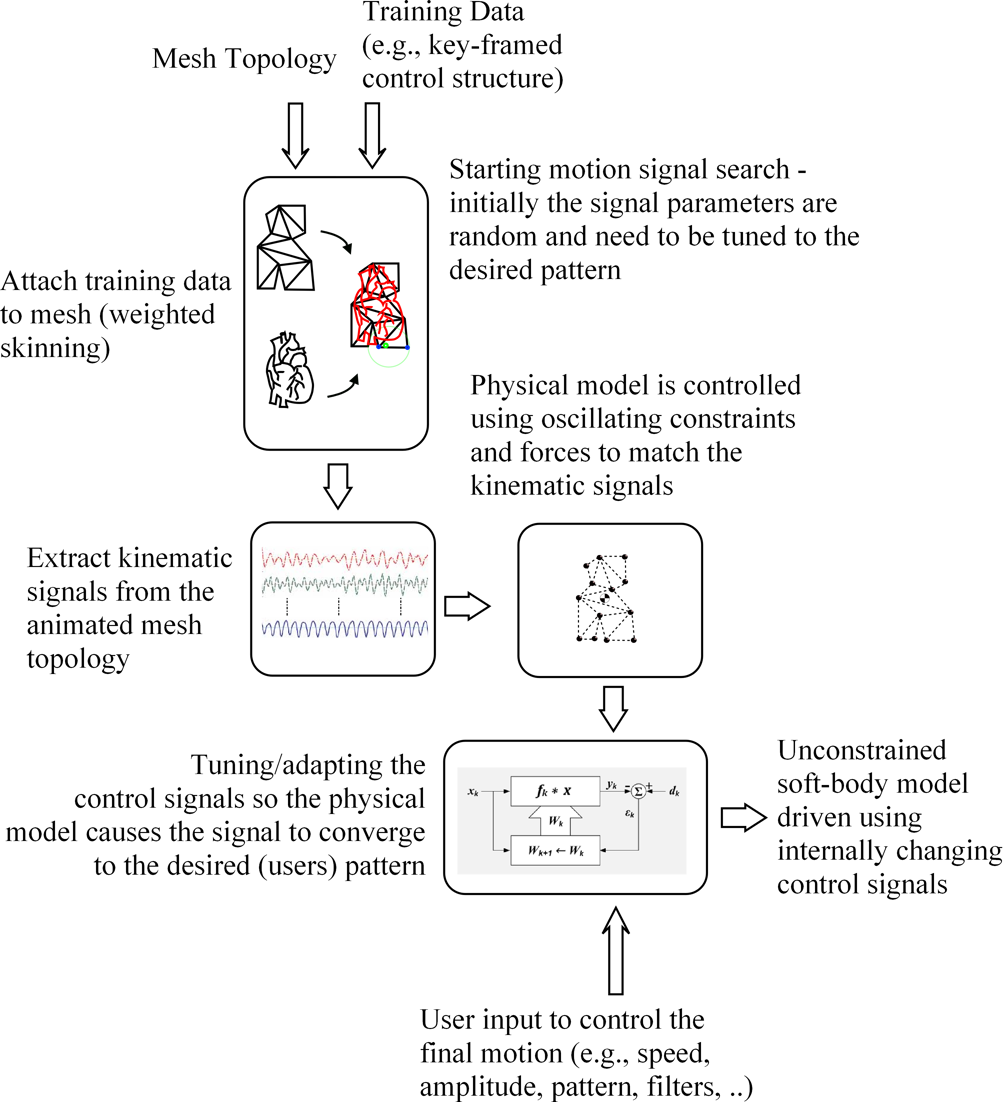}
{Overview}
{Interconnected elements to construct and tune the soft-body's self-driven motions.}
{1.0}

\paragraph{Visualization}
The graphical simulation of internal self-driven organs assists in visualizing and understanding the function of complex systems.
% learning fulfils an important need for pictorial representation of the functions of organs and systems. 
We integrate a physically-based computer animation technique with low-dimensional medical models.
The generation and customization of active muscle animations is a resource-intensive problem. 
Rough data sets are available for providing outline motion characteristics (i.e., overall changes of size and shape of the organ from ultrasound scans and medical x-rays to help train the initial animation parameters).
%
% Images can be drawn as originals or can be copiedhcanned from various sources. 
% By 
This rough motion data provides a simple and approximate initial (starting) pattern for the training algorithm.  
% to the particular/basic need of the teacher and projecting the end-point image by using a vector animation package, 
Eventually, the overall animation converges on an oscillatory pattern over time as the system settles. % 'films' can be created to demonstrate any form of movement.
% In the Anatomy Department, Sultan Qaboos University in Muscat, computer-animated tutorials are being introduced to illustrate normal and abnormal functional anatomy. The heart and its valve mechanisms have been selected as a pilot study. The student response is very positive and the technique has great potential. Embryology animations showing the formation and growth of organs such as the brain and spinal cord are also being developed.
Depending on the tuned parameters, the final result can positive or negative.  Since the mechanistic motion can converge on abnormal patterns.  Of course, this could be used to identify and illustrate `incorrect' organ traits - aiding students to by showing them correct and incorrect motions patterns.

\paragraph{Limitations}
We focused on individual organs working alone.
Lacked the coupled inter-organ interaction (coupled influence).
Not to mention, the simulation system did not embrace any fluid mechanic principles that would be inherent in a real-world case (e.g., organs would be immersed in flowing blood).
Our lightweight model exploited a low-dimensional view as not to hinder performance, while still able to capture the rhythmic organic properties of the organs motion (but driven by physical-laws, tunable and interactive).

\figuremacroWb
{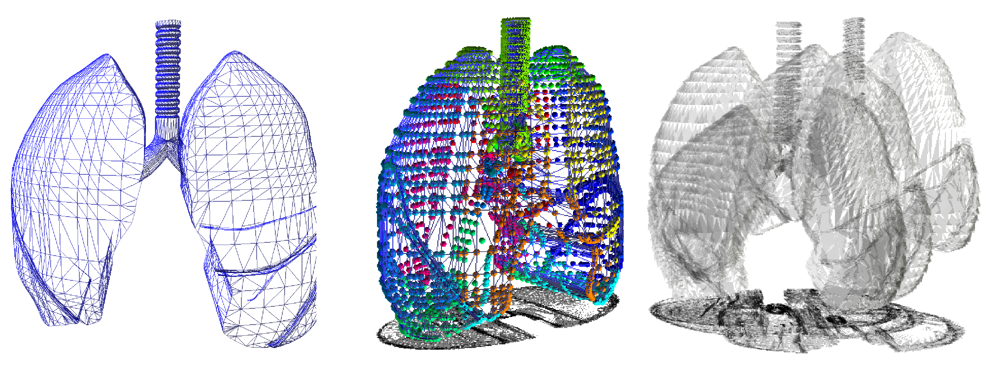}
{Lungs}
{Three-dimensional mesh model of the organ (e.g., lung), classify different regions of the mesh (small sub-areas), oscillating motion based on the contraction/expansion of tuned constraints.}
{1.0}

\section{Experimental Results} \label{sec:results}
For experimentation, we imported three-dimensional organ meshes.
We broke the overall mesh into different areas that could be classified with custom parameters.  For instance, the organ valves (input/output) would contract/oscillate less (more rigid and fixed) (see Figure \ref{fig:heart} and Figure \ref{fig:lungs}).
We only experimented with aesthetic meshes, that is, we did not create detailed representations with internals. %  tissue workings.
For instance, the heart has internal structures, such as, the semilunar valve and chordae tendineae - yet we did not simulate flow, pressure or internal contacts/collisions so did not incorporate them into the simulation.

\figuremacroW
{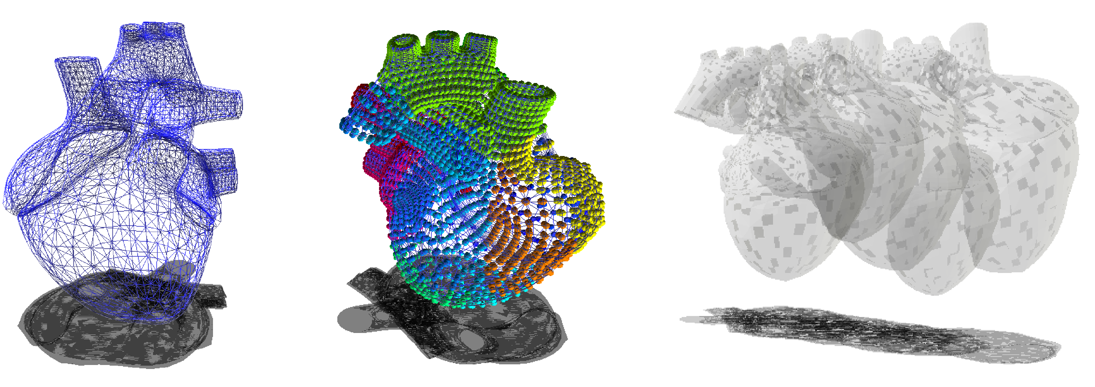}
{Heart}
{Three-dimensional mesh model of the organ (e.g., heart), classify different regions of the mesh (small sub-areas), oscillating motions based on the contraction/expansion of tuned constraints.}
{1.0}

\section{Conclusion and Discussion} \label{sec:conclusion_and_discussion}
We presented a self-driven soft-body technique for emulating the internal rhythmic motion of organs. 
Although many techniques have been proposed for simulating deformable
organic tissue \cite{webster2002elastically}, few provide tunable motion control for emulating internal muscle operations.
%) the  performance necessary for
%real-time applications. 
Our scalable model provides trade-off's between resolution accuracy and interactive performance - which is important for real-time solutions.
Furthermore, since our solution is based on a physical model the simulation is able to incorporate and feedback forces from external contacts.
The work has only scratched the potential, as there are a whole number of facts and possibilities to investigate, for example, when we listen to music, our heartbeat signal changes to mimic the music \cite{lundqvist2008emotional}.
In essence, our early solution provides a promising visualization tool for medical educational and training.

Various muscles within the heart oscillate with different frequencies.
The inter-collaboration of these frequencies is important so the heart is able perform optimally.  While our work used a simplified three-dimensional model - further investigations would be to increase the complexity of the anatomical model and perform further training/analysis to determine/understand possible signal patterns (for training and analysis).

\figuremacroW
{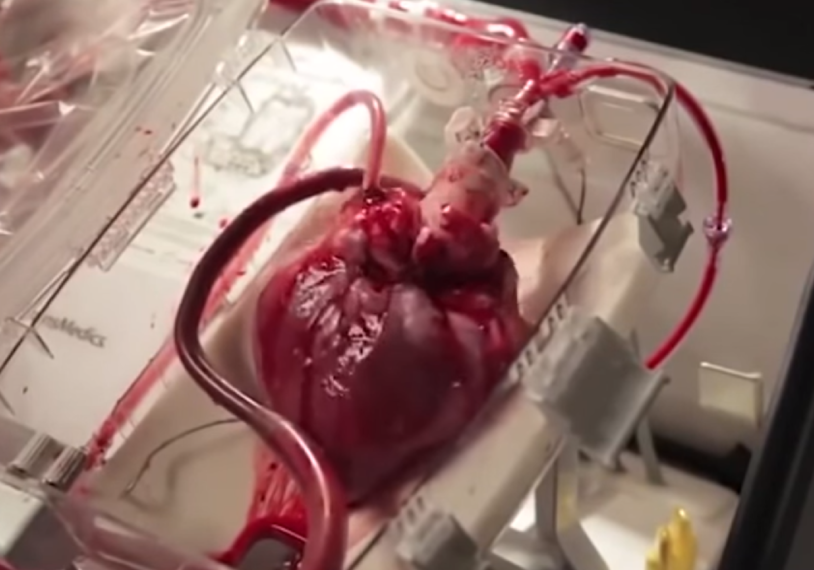}
{Heart Beating Outside Body}
{The heart is a complex self-driven organ (muscle) that is able to beats outside of the body \cite{robicsek1963simple}.}
{1.0}

\bibliographystyle{ieeetr}

\bibliography{paper} % bib filename

\figuremacroW
{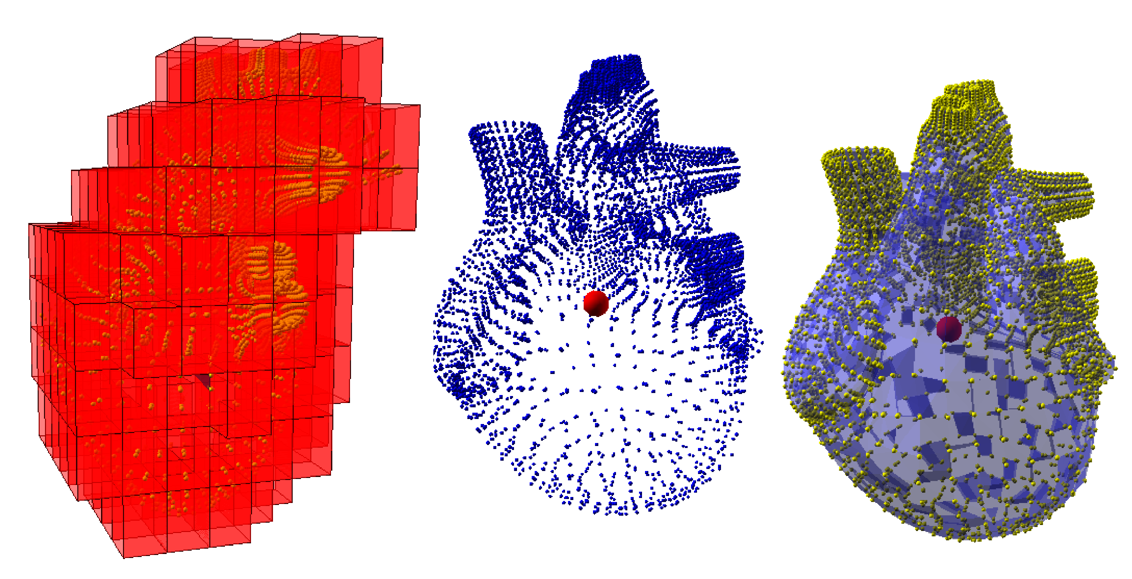}
{Regionalising}
{Reducing the mesh complexity by partition up the model into regions.  Providing clusters of vertices compared to solving each inter-particle constraints - problem more scalable and viable for real-time experimentation (regions are flexible - with the ability to add more or less to particular areas of importance).}
{1.0}

\end{document}